\documentclass{aa}  

\usepackage{dashbox}
\usepackage{framed,color,ocg-p}
\newcommand{\ToggleLayer}[2]{%
  \leavevmode
  \pdfstartlink user {
    /Subtype /Link
    /Border [0 0 0]%
    /A <<
      /S/JavaScript
      /JS (
         var aOCGs = this.getOCGs(), Layer;
         var Layers = "#1".split(","), Active = -1, i, l;
         for (l=0; l<Layers.length; l++) {
           Layer = Layers[l];
           for (i=0; aOCGs && i<aOCGs.length; i++) {
             if (aOCGs[i].state && aOCGs[i].name == Layer) {
               Active = l;
               aOCGs[i].state = false;
             }
           }
           if (Active >= 0) break;
         }
         if (Active == -1) {
           for (l=0; l<Layers.length; l++) {
             if (Layers[l] == "") Active = l;
           }
         }
         Active = Active + 1;
         if (Active == Layers.length) Active = 0;
         Layer = Layers[Active];
         for (i=0; aOCGs && i<aOCGs.length; i++) {
           if (aOCGs[i].name == Layer) aOCGs[i].state = true;
         }
      )
    >>
  }#2%
  \pdfendlink
}

\usepackage{graphicx}
\usepackage{txfonts}
\newcommand{\diff}{\mathrm d}
\begin{document}

\title{2MASS wide-field extinction maps: V. Corona Australis}
\author{Jo\~ao Alves\inst{1}, Marco Lombardi\inst{2}, and Charles
  J. Lada\inst{3}}
\institute{University of Vienna, Department of Astrophysics,
  T\"urkenschanzstrasse 17, 1180 Vienna, Austria
\and 
University of Milan, Department of Physics, via Celoria 16, I-20133
Milan, Italy 
\and
Harvard-Smithsonian Center for Astrophysics, Mail Stop 42, 60 Garden
Street, Cambridge, MA 02138, USA}
\date{Received June 2013; Accepted ***date***}

 
\abstract{ 
We present a near-infrared extinction map of a large region ($\sim$870 deg$^2$) covering the isolated Corona Australis complex of molecular clouds.  
We reach a 1-$\sigma$ error of 0.02 mag in the K-band extinction with a resolution of 3 arcmin over the entire map. We find that the Corona Australis cloud is about three times as large as revealed by previous CO and dust emission surveys. The cloud consists of a 45 pc long complex of filamentary structure from the well known star forming Western-end (the head, $N \geq10^{23}$ cm$^{-2}$) to the diffuse Eastern-end the tail, ($N \leq10^{21}$ cm$^{-2}$). Remarkably, about two thirds of the complex both in size and mass lie beneath A$_V\sim1$ mag.  We find that the probability density function (PDF) of the cloud cannot be described by a single log-normal function. Similar to prior studies, we found a significant excess at high column densities, but a log-normal + power-law tail fit does not work well at low column densities.  We show that at low column densities near the peak of the observed PDF, both the amplitude and shape of the PDF are dominated by noise in the extinction measurements making it impractical to derive the intrinsic cloud PDF below A$_K <$ 0.15 mag. Above A$_K \sim 0.15$ mag, essentially the molecular component of the cloud, the PDF appears to be best described by a power-law with index $-3$, but could also described as the tail of a broad and relatively low amplitude, log-normal PDF that peaks at very low column densities.
}

 \keywords{ISM: clouds, dust, extinction, ISM: structure, ISM:
    individual objects: Corona Australis molecular complex, Methods: data analysis}

   \maketitle
%

\section{Introduction}
\label{sec:introduction}

This paper is the fifth in a series where we apply an optimized multi-band technique dubbed Near-Infrared Color Excess Revisited (\textsc{Nicer}, \citealp{Lombardi2001}, hereafter Paper~0) to measure dust extinction and study the structure of nearby molecular dark clouds using the Two Micron All Sky Survey (2MASS, \citealp{Kleinmann1994}). Compared to all-sky extinction mapping, dedicated studies of complexes have the advantage of 1) having optimal zero point calibrations from a careful choice of a local control field, and 2) optimal resolution, tuned to the local background stellar density.  Previously, we studied the Pipe nebula (see \citealp{Lombardi2006}, hereafter Paper I), the Ophiuchus and Lupus complexes (\citealp{Lombardi2008}, hereafter Paper II), the Taurus, Perseus, and California complexes (\citealp{Lombardi2010}, hereafter paper III), and the Orion, Rosette, and CMa complexes \citep{2011A&A...535A..16L}.  In this paper we present a wide-field extinction map, constructed from 10.7 million stars from the 2MASS database, of a large region covering $\sim$ 870 deg$^2$ centered on the Corona Australis complex. 

The main aim of our coordinated study of nearby molecular clouds is to investigate in detail the large-scale structure of these clouds, down to the lowest column densities measurable with this technique, which are often below the column density threshold required for the detection of the CO molecule \citep[e.g.][]{Alves1999,Lombardi2006}. In addition, the use of an uniform dataset and of a consistent and well tested pipeline allows us to characterize many properties of molecular clouds and to identify cloud-to-cloud variations in such properties.

The advantages of using near-infrared dust extinction as a column density tracer have been discussed elsewhere \citep{Lada1994,Alves1999,Lombardi2001,Lombardi2006}. Recently, \citet{Goodman2009} used data from the COMPLETE survey \citep{Ridge2006} to assess and compare three methods for measuring column density in molecular clouds, namely, near-infrared extinction (\textsc{Nicer}), dust thermal emission in the mm and far-IR, and molecular line emission. They found that observations of dust are a better column density tracer than observations of molecular gas (CO), and that observations of dust extinction in particular provide more robust measurements of column density than observations of dust emission, mainly because of the dependence of the latter measurements on uncertain knowledge of dust temperatures and emissivities.

This paper uses, in some key analyses, the improved \textsc{Nicest} method \citep{Lombardi2009}, which copes well with the sub-pixel inhomogeneities present in the high-column density regions of the maps. The inhomogeneities can either be due to the vanishing number of stars towards these regions of the map, to steep gradients in the column density map, to the effect of turbulent fragmentation, or to the increased presence of foreground stars, and they bias the measurements towards lower column densities. All these effects are expected to be most severe in the densest regions of dark complexes, i.e. in a very small fraction of the large areas considered in this paper. Nevertheless, because of the relevance of these regions in the process of star formation, it is important to understand this bias and how to correct for it, as is done in the \textsc{Nicest} method.

\begin{figure}[!tbp]
  \centering
  \begin{ocg}{fig:0a}{fig:0a}{0}%
    \includegraphics[width=\hsize]{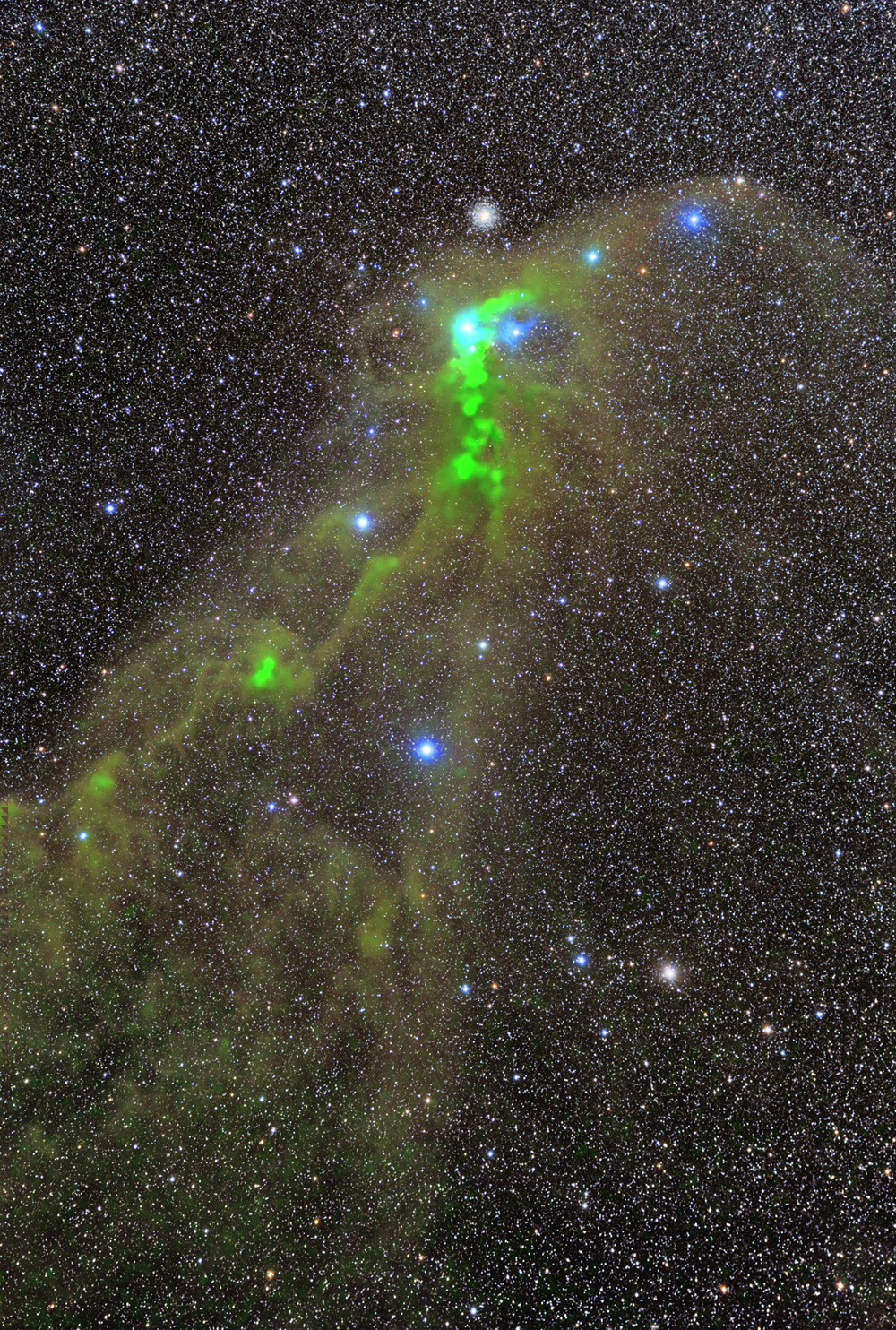}%
  \end{ocg}%
  \hspace{-\hsize}%
  \begin{ocg}{fig:0b}{fig:0b}{1}%
    \includegraphics[width=\hsize]{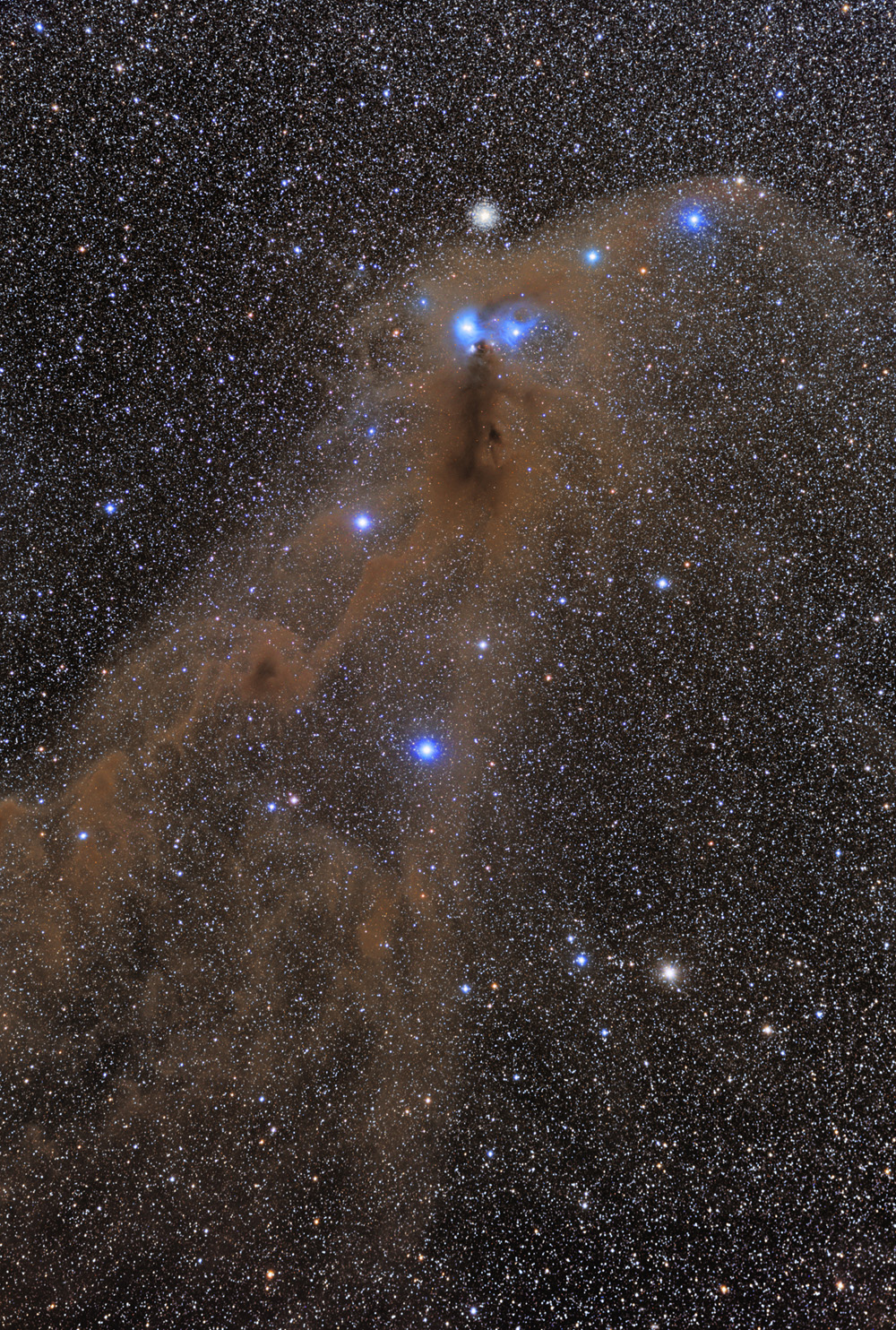}
  \end{ocg}%
  \caption{Optical image of the Corona Australis complex. The image
    covers approximately 3 by 5 deg$^2$ and covers the Western-end of
    the cloud, the most dense region of the cloud where star formation
    is ongoing. By clicking the toggle image box below while using
    Acrobat Reader, the extinction map presented in this paper appears
    in green. Image courtesy of Pavel Pech.
    \ToggleLayer{fig:0a,fig:0b}{\protect\dbox{Toggle image}}}
  \label{fig:0}
\end{figure}

Corona Australis (see Figure~\ref{fig:0}) is one of the most efficient star forming regions \citep{Lada2010}, is located at a distance of about 130 pc \citep{Casey1998}, and harbors an embedded cluster towards the Western-end of the cloud (see Figure~\ref{fig:0}) with about 50 young stars (\citet{Forbrich2007}; for a recent review of the region see \citealp{Neuhauser2008}). The large scale structure of the denser regions of the complex was studied in C$^{18}$O by \citet{Harju1993} and in dust emission by \citet{Chini2003}.  Recently, \citet{Peterson2011} presented Spitzer IRAC and MIPS observations of a 0.85 deg$^2$ field centered on the Corona Australis star-forming region, on the Western-end of the cloud (Figure~\ref{fig:0}). Combining the Spitzer results and data form the literature these authors find a total of 116 candidate young stellar objects (YSOs) and further evidence that star formation is ongoing in the cloud. Using high-resolution spectroscopy from the VLT for a sample of 18 YSOs in the Coronet, \citet{Sicilia-Aguilar2011} determines an age of $<2$ Myr, and probably $\sim0.5-1$ Myr for the age of this cluster. An extinction map of the $\sim3^\circ\times6^\circ$ densest regions of the Corona Australis, again the Western-end of the cloud, appeared in \citet{Schneider2011}, as part of a multi-tracer study of a sample of clouds that was used to quantify a possible link between cloud structure and turbulence.

One of the major complications in studying molecular cloud structure is to account for the possibility that two or more clouds are seen along the same line-of-sight. When this is the case, even if one of the clouds is only a diffuse cloud, the structure analysis is further complicated (see for example the probability distribution of pixel extinctions in the Pipe Nebula, Figure 20 of \citet{Lombardi2006}). This complication is minimized and can be safely ignored for relatively high Galactic-latitude clouds like Corona Australis that lies at a projected distance from the Galactic plane of $\sim20^{\circ}$. Corona Australis is one of the most isolated Galactic star forming cloud as seen from Earth which makes it an ideal case study for molecular cloud structure and star formation.

This paper is organized as follows. In Sect.~\ref{sec:nicer-absorpt-map} we briefly describe the technique used to map the dust and we present the main results obtained. A statistical analysis of our results and a discussion of the bias introduced by foreground stars and unresolved substructures is presented in Sect. 3. Section 4 is devoted to the mass estimate of the cloud complexes. Finally, we summarize the results obtained in this paper in Sect. 5.

\section{\textsc{Nicer} and \textsc{Nicest} extinction maps}
\label{sec:nicer-absorpt-map}

The data analysis was carried out following the \textsc{Nicer} and \textsc{Nicest} techniques and used also in the previous papers of this series, to which we refer for the details (see in particular Paper~III).  We selected reliable point source detections from the Two Micron All Sky Survey\footnote{See \texttt{http://www.ipac.caltech.edu/2mass/}.} (2MASS, \citet{Kleinmann1994}) in the region:

\begin{align}
  \label{eq:1}
  -20^\circ <{} & l < 20^\circ \; , & -37^\circ <{} & b < -13^\circ \; .
\end{align}

This area ($\sim870$ deg$^2$ containing approximately $10.7$ million point sources from the 2MASS catalog) contains the Corona Australis cloud complex and its mainly dust free environment. 

\begin{figure*}[!tbp]
  \begin{center}
\includegraphics[width=\hsize]{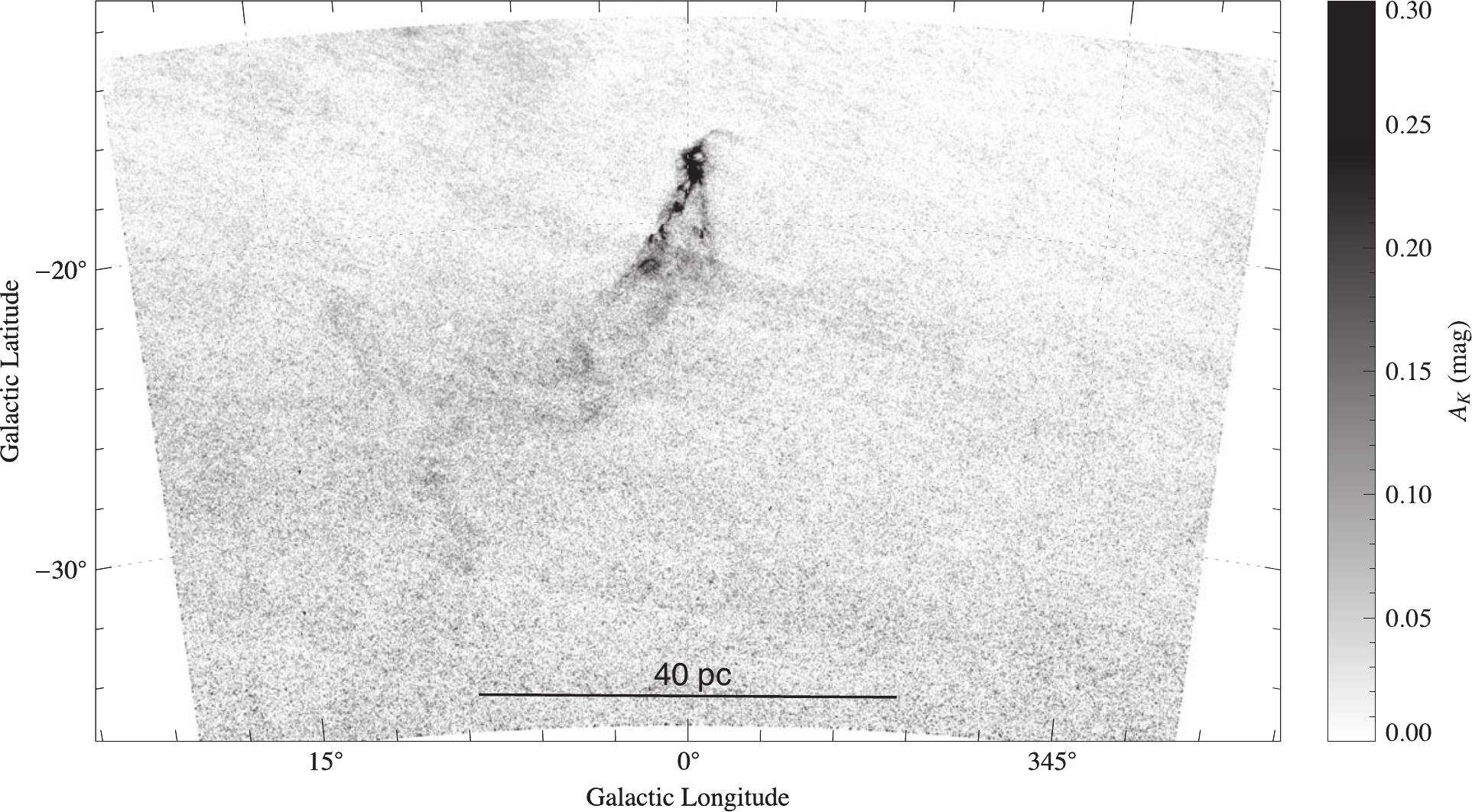}
\caption{The \textsc{Nicer} extinction map of the Corona Australis
  cloud complex.  The resolution of the map is $\mathrm{FWHM} = 3
  \mbox{ arcmin}$ and covers an area of $\sim$ 870 deg$^2$. About 10.7
  million stars from the 2MASS database were used in the construction
  of this map.  This figure suggests an empirical division of two structurally different cloud components: 1) a ``head'' comprising the denser (A$_K > 0.1$ mag), highly structured filament and 2) a ``tail'' of more diffuse, less structured cloud material.}
    \label{fig:3}
  \end{center}
\end{figure*}

\begin{figure*}[!tbp]
  \begin{center}
\includegraphics[width=\hsize]{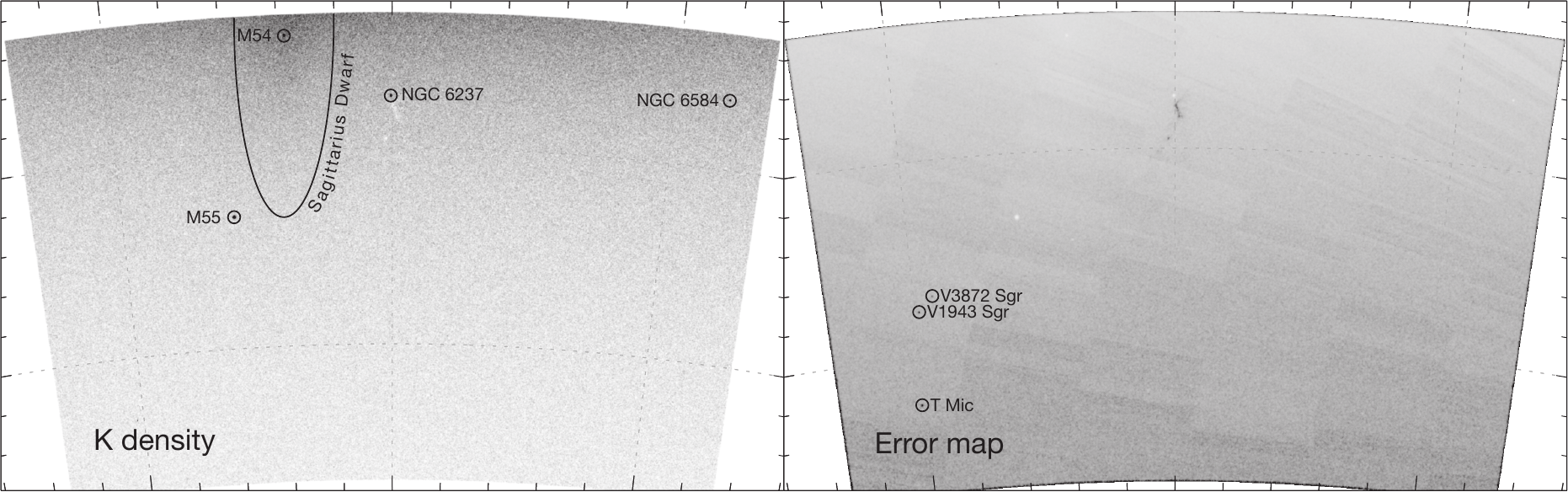}
\caption{Left: The K-band stellar density of the entire field studied
  (same scale as in Figure \ref{fig:3}). Clusters in this image appear as point
  sources and are labeled. The half ellipse marks an extended stellar
  overdensity corresponding to the Sagittarius Dwarf galaxy. Right:
  Error map of the \textsc{Nicer} extinction map. The average error in the map
  is A$_K$$\sim$0.02 mag. The marked error peaks are caused by very
  bright, saturated, red giants that create a no-data region. The
  maximum error towards the densest regions of the cloud is
  A$_K$$\sim$0.06 mag.}
    \label{fig:2}
  \end{center}
\end{figure*}

\begin{figure*}[!tbp]
  \begin{center}
\includegraphics[width=\hsize]{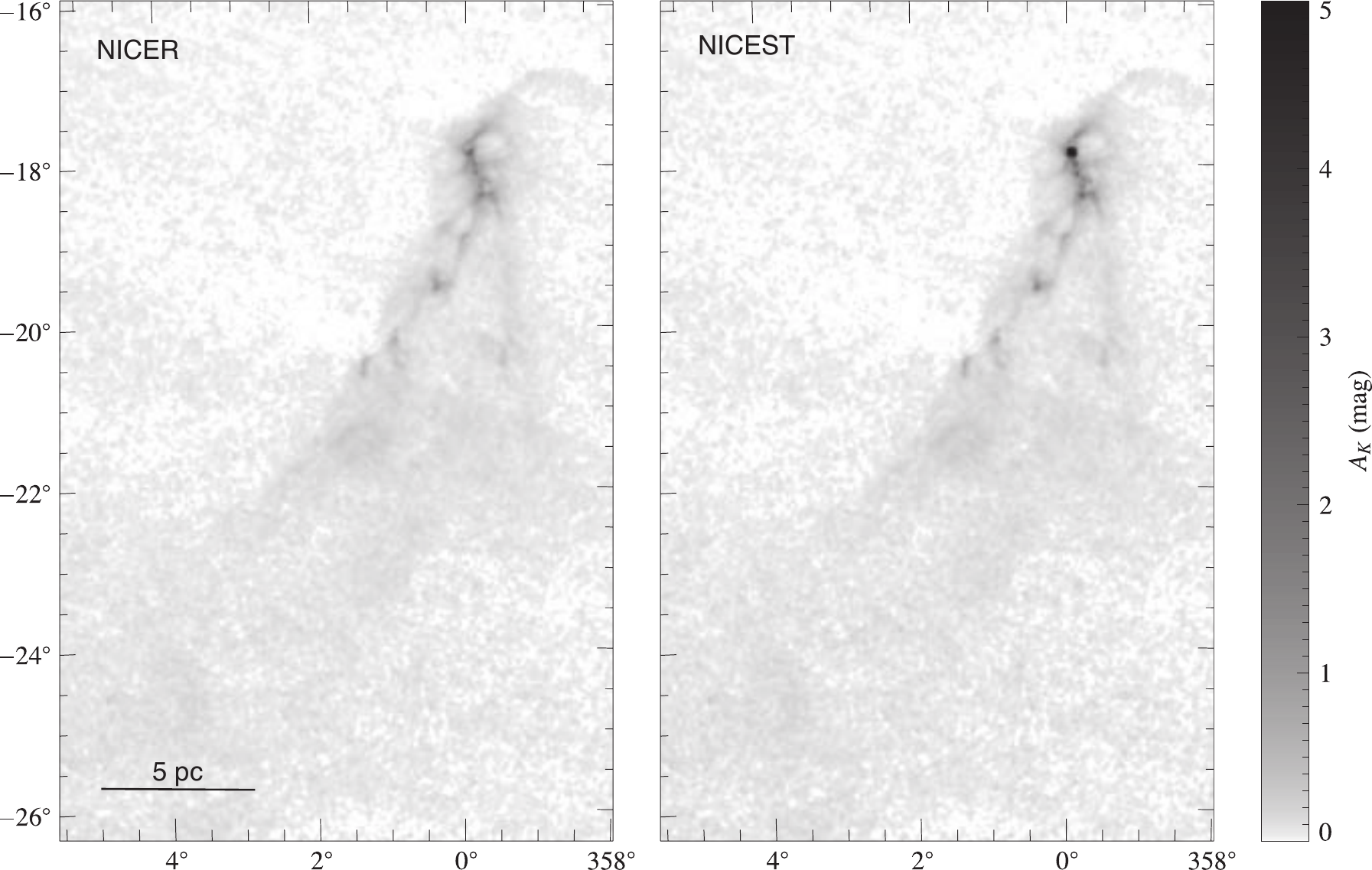}
\caption{Comparison between the \textsc{Nicer} map (left) and the
  \textsc{Nicest} map (right) of the Corona Australis cloud
  complex. The grayscale is the same for both maps. }
    \label{fig:4}
  \end{center}
\end{figure*}

As a preliminary check, we considered the color-color diagram of the stars selected to verify the possible presence of obvious anomalies in the extinction law. Unlike Paper~II, we find only a weak sign of possible contamination by evolved stars and decided to  proceed similarly to Paper~III by retaining all objects.

After the selection of a control field for the calibration of the intrinsic colors of stars (and their covariance matrix) we produced the final 2MASS/\textsc{Nicer} extinction map, shown in Figure~\ref{fig:3}. The selected control field was defined as  a circle of $\sim 4.5^\circ$ centered on $l=349^\circ$ and $b=-17^\circ$. For best results, we smoothed the individual extinctions measured for each star, $\bigl\{ \hat A^{(n)}_K \bigr\}$, using a moving weight average \begin{equation}
  \label{eq:2}
  \hat A_K(\vec\theta) = \frac{\sum_{n=1}^N W^{(n)}(\vec\theta) \hat
    A^{(n)}_K }{\sum_{n=1}^N W^{(n)}} \; ,
\end{equation}
where $\hat A_K(\vec\theta)$ is the extinction at the angular position $\vec\theta$ and $W^{(n)}(\vec\theta)$ is the weight for the $n$-th star for the pixel at the location $\vec\theta$.  This weight, in the standard \textsc{Nicer} algorithm, is a combination of a smoothing, window function $W\bigl( \vec\theta - \vec\theta^{(n)} \bigr)$, i.e.\ a function of the angular distance between the star and the point $\vec\theta$ where the extinction has to be interpolated, and the inverse of the inferred variance on the estimate of $A_K$ from the star:
\begin{equation}
  \label{eq:3}
  W^{(n)}(\vec\theta) = \frac{W \bigl( \vec\theta - \vec\theta^{(n)}
    \bigr)}{\mathrm{Var}\bigl( \hat A^{(n)}_K \bigr)} \; .
\end{equation}
For this paper, the weight function $W$ was taken to be a Gaussian with $\mathit{FWHM} = 3 \mbox{ arcmin}$.  Finally, the map described by Eq.~\eqref{eq:2} was sampled at $1.5 \mbox{ arcmin}$ (corresponding to a Nyquist frequency for the chosen weight function).

The \textsc{Nicer} extinction map in presented in Figure~\ref{fig:3}. The first striking impression from the map is the size of the cloud, extending across the sky for about 45 pc. Existing molecular line and dust emission studies \citep[e.g.][]{Harju1993,Dame2001,Chini2003} have revealed an extension about three times smaller, coinciding with the denser filamentary structure approximately between $-17^\circ > l > -22^\circ $.  Not surprisingly, this is the region of the cloud where column density is above A$_K \sim 0.1$ mag (or A$_V \sim 1$ mag), which is close to the threshold for the formation and detection of CO \citep[e.g.][]{Alves1999,Lombardi2006,Pineda2010}. Figure~\ref{fig:3} suggests an empirical division of two structurally different cloud components: 1) a ``head'' comprising the denser (A$_K > 0.1$ mag), highly structured filament at  $l > -22^\circ $, detected by previous CO surveys and 2) a ``tail'' of more diffuse, less structured cloud material, about twice as large as the ``head'' and in all likelihood with no CO emission (A$_K < 0.1$ mag, $l < -22^\circ $). 

In Figure~\ref{fig:2} we present the K-band stellar density map (left) and the error map (right) for the entire region used to construct the extinction map. The K-band stellar density map shows, as expected, a smooth gradient across Galactic latitude, but also point-like sources corresponding to globular clusters. Less obvious but clearly present is an extended overdensity corresponding to the Sagittarius Dwarf galaxy.  The extinction error map (right) was evaluated from a standard error propagation in Eq.~\eqref{eq:2} (see Paper~I).  The main factor affecting the error is the local density of stars, and thus there is a detectable gradient in the statistical error also across Galactic latitude.  Other variations can be associated with bright stars (which are masked out in the 2MASS data base, creating no-data regions), clusters (expectedly creating low error regions), bright galaxies, and to the cloud itself.  All the point-like error peaks in the error map correspond to (saturated) red giants. For reference, the brightest saturated red giant T Mic has an apparent magnitude on the K-band of $\sim-1.6$.  The average error shown in Figure~\ref{fig:2} is $0.02 \mbox{ mag}$.

Similarly to Paper~III, we also constructed a \textsc{Nicest} extinction map, obtained by using the modified estimator described in \citet{Lombardi2009}.  The \textsc{Nicest} map differs significantly from the \textsc{Nicer} map only in the high column-density regions, where unresolved cloud structures produce a possibly significant bias on the standard estimate of the column density.  As an example we present in Figure \ref{fig:4},  a comparison between the \textsc{Nicer} and the \textsc{Nicest} extinction map of the ``head'' of the cloud. As it can be seen, the only difference between both maps occurs for the areas of high column density, where \textsc{Nicest} is able to correct for the bias introduced by the fact that there are few background stars detected at the highest extinctions. Indeed, the largest extinction in both maps was measured towards the Coronet cluster, coinciding with the brightest dust emission peak in \citet{Chini2003}, namely MMS13, where $A_K \simeq 2.4$ magnitudes for \textsc{Nicer} and $A_K \simeq 5.4$ magnitudes for \textsc{Nicest}.

 \section{Results}
\label{sec:statistical-analysis}

\subsection{Reddening law}
\label{sec:reddening-law}

\begin{figure}[!tbp]
  \begin{center}
    \includegraphics[width=\hsize]{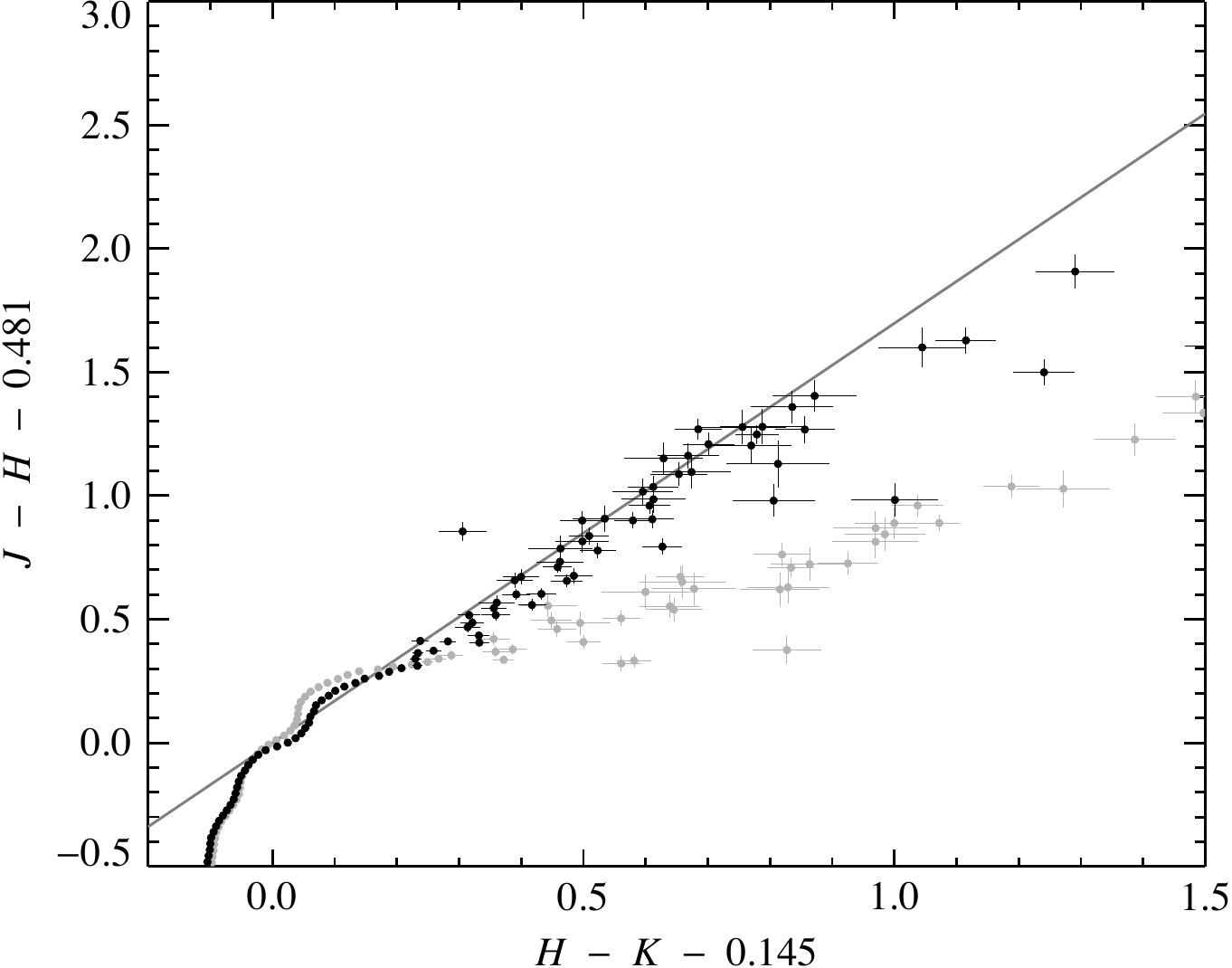}
    \caption{Verification of the reddening law.  The points show the
      color excess E($J - K$) as a function of the color excess E($H -
      K$). The gray measurements are contaminated by the Sagitarius
      Dwarf galaxy (see Figure~\ref{fig:2}).  The solid line shows the
      normal infrared reddening law \citep{Indebetouw2005}.  Error
      bars are uncertainties evaluated from the photometric errors of
      the 2MASS catalog. }
    \label{fig:8}
  \end{center}
\end{figure}

As shown in \citet{Lombardi2001}, the use of multiple colors in the \textsc{Nicer} algorithm significantly improves the signal-to-noise ratio of the final extinction maps, but also provides a simple, direct way to verify the reddening law.  A good way to perform this check is to divide all stars with reliable measurements in all bands into different bins corresponding to the individual $\hat A_K^{(n)}$ measurements, and to evaluate the average NIR colors of the stars in the same bin. Figure~\ref{fig:8} shows the results obtained for a bin size of $0.02$ mag, the average error in our extinction map.

Unlike the previous papers of the series, where there is in general a good correlation between both quantities, the distribution of data points for Corona Australis is non trivial: while some points seem to follow the normal infrared reddening law, others consistently deviate from it. The deviant sequence of points seem to suggest either 1) a region of the cloud with a shallower reddening law or 2) the presence of a different, intrinsically redder, stellar background population. To investigate the latter, in particular to verify the effect of the Sagitarius Dwarf on the background population, we divided the image at about $l=3^\circ$ and plotted  measurements lying at $l > 3^\circ$ in gray (the area containing the dwarf galaxy), and in black the measurements lying at $l < 3^\circ$ (the area containing most of the complex). Inspection of Figure~\ref{fig:8} suggests strongly that the deviant sequence is likely caused by the presence of an intrinsic redder stellar population belonging to the Sagittarius Dwarf. 

\begin{figure}[!tbp]
  \begin{center}
    \includegraphics[width=\hsize]{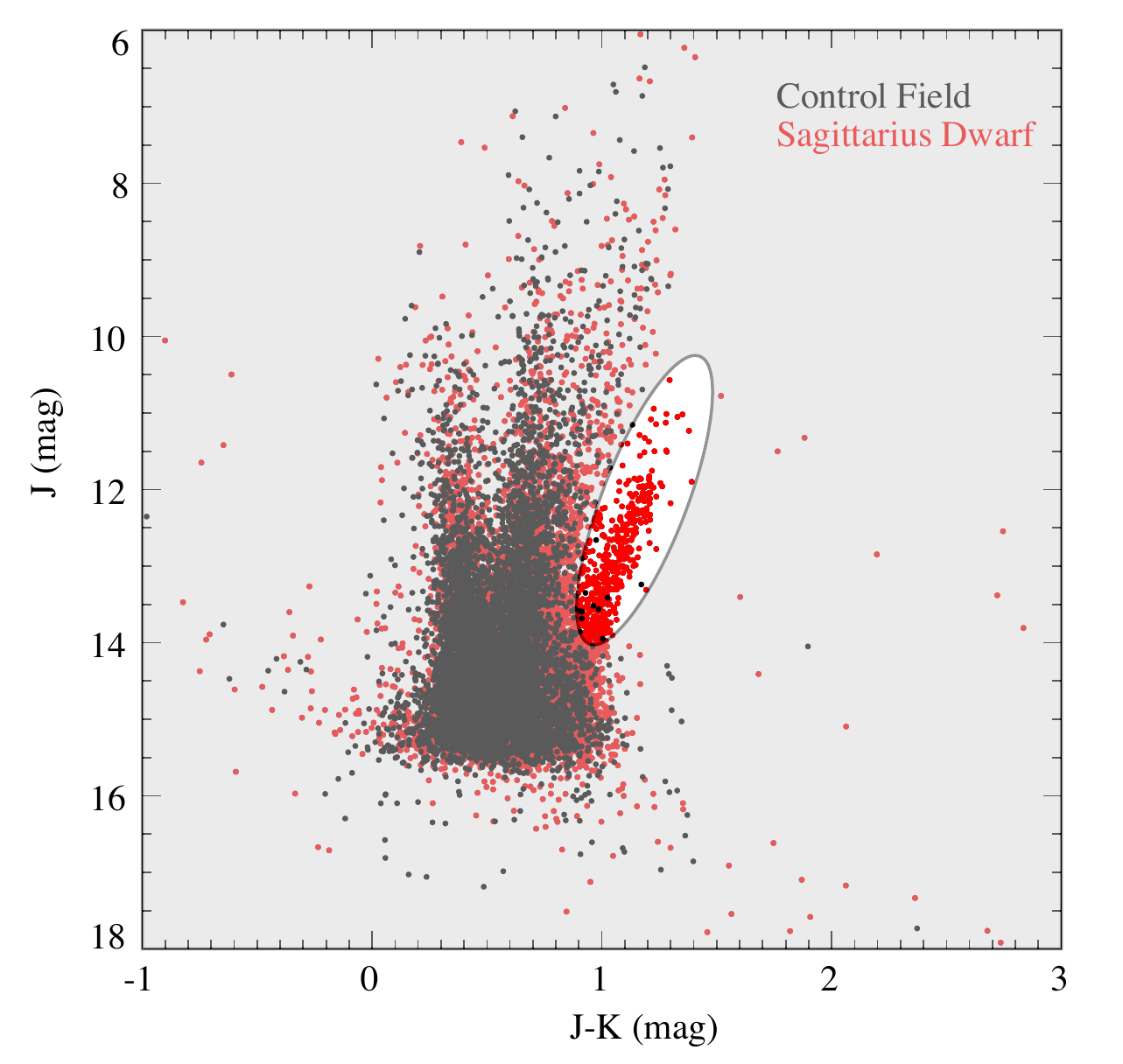}
    \caption{Color-Magnitude diagram for a 1 degree diameter field towards the Sagittarius Dwarf Galaxy ($l=6^\circ, b=-14^\circ$) and for an equal size Control Field taken at the same Galactic latitude ($l=351^\circ, b=-14^\circ$). }
    \label{fig:SDG_onoff}
  \end{center}
\end{figure}

\begin{figure*}[!tbp]
  \begin{center}
    \includegraphics[width=\hsize]{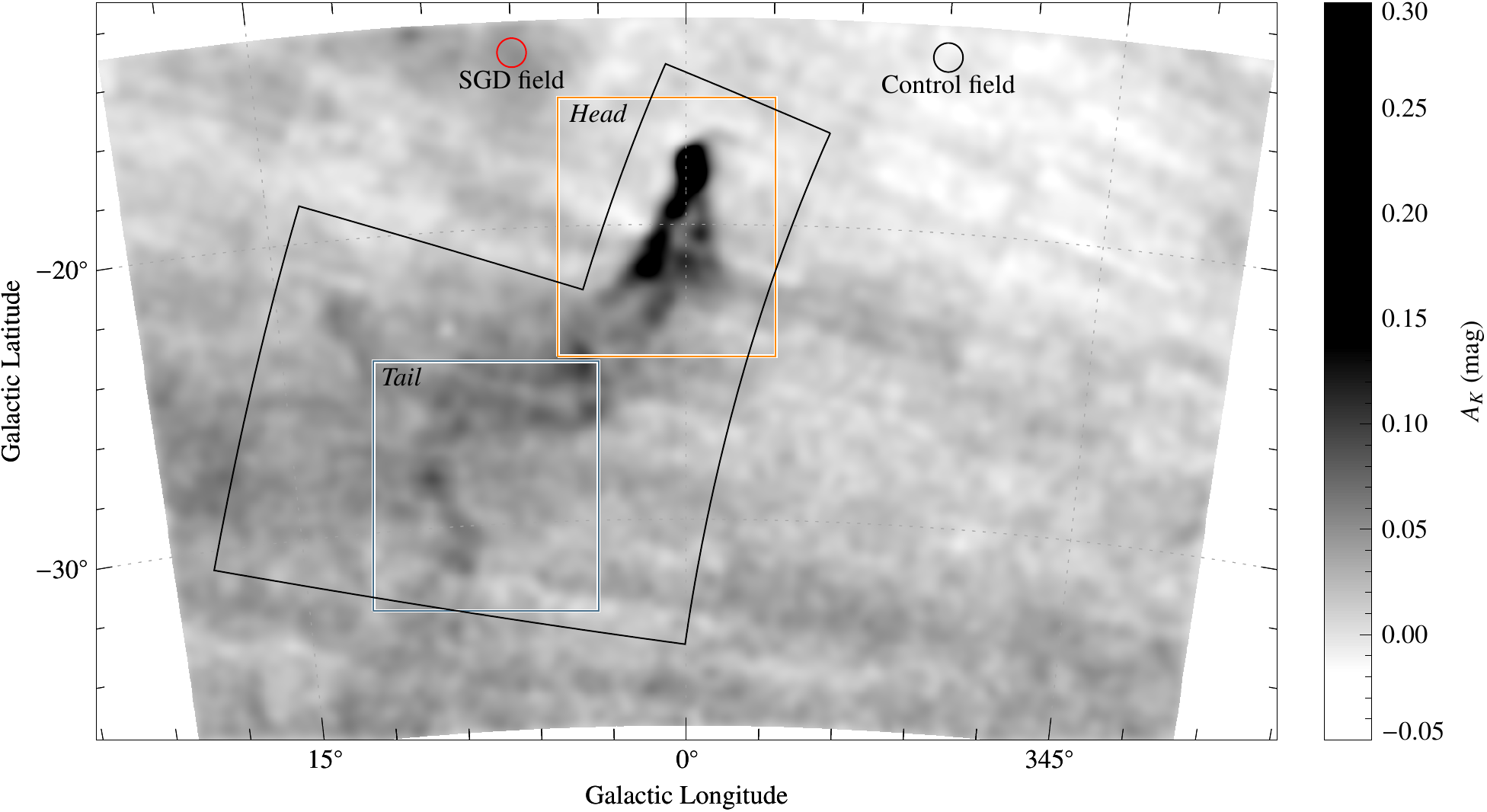}
    \caption{The extinction map of the Corona Australis cloud complex
      smoothed with a gaussian kernel.  The resolution of the map is
      $\mathrm{FWHM} = 1 \mbox{ deg}$. The area enclosed by the solid
      polygon contains the cloud complex while avoiding contamination
      from the Sagitarius Dwarf Galaxy (SDG) and was used to compute
      the mass of the cloud (in Section~\ref{sec:mass-estimate}). The
      two circles delimit the SDG field (red) and the Control field
      (black) discussed in Figure~\ref{fig:SDG_onoff}. The
      \textit{``head''} (orange) and the \textit{``tail''} (blue)
      regions are discussed in Section~\ref{sec:column-dens-prob}.}
    \label{fig:5}
  \end{center}
\end{figure*}

To investigate this further we constructed a $J-K$ vs. $J$ color-magnitude diagram (Figure~\ref{fig:SDG_onoff}) of a circular 1 degree diameter field centered on the Sagittarius Dwarf ($l=6^\circ, b=-14^\circ$) and an equal size Control Field taken at the same Galactic latitude ($l=351^\circ, b=-14^\circ$), but 15$^\circ$ away in galactic longitude (see Figure~\ref{fig:5}). It immediately becomes clear that the Sagittarius Dwarf field contains a third and redder branch that is not present in the control field. This branch is likely populated by M-giants belonging to the Sagittarius Dwarf \citep[e.g.][]{Majewski2003}, and it is widely used to trace the extent of the Dwarf galaxy and its tidal tail because of its relative intrinsic brightness and distinct NIR color from the Milkyway field. But exactly because these M-giants are see only towards the Sagittarius Dwarf, they are not accounted for in the characterization of the unreddened stellar background field to this cloud and so they will artificially increase the extinction measurement towards their direction.

The overlap between the galaxy and the cloud is not severe, luckily, so we can attempt to confine the region of the map that is affected by the redder Sagittarius Dwarf stellar population, and not use it to determine cloud parameters. Still, the Sagittarius Dwarf Galaxy has a long tidal tail, so some overlap is to be expected.

Finally, we stress that we are not deriving the reddening law, but only checking its consistency with respect to a standard law. The check performed with Figure~\ref{fig:8} is a \textit{relative\/} one: we can only verify that the extinction derived from the $H - K$ and $J - H$ colors agree with expectations from the standard \citep{Indebetouw2005} reddening law.  The discrepancy observed for the less contaminated sample (black points) is relatively small and is likely to produce a systematic bias on the final column density of our maps below $5\%$.

\subsection{Mass estimate}
\label{sec:mass-estimate}

In order to obtain the Corona Australis cloud mass $M$, we use the standard relation 

\begin{equation}
  \label{eq:14}
  M = d^2 \mu \beta_K \int_\Omega A_K(\vec\theta) \, \diff^2 \theta \; ,
\end{equation}

where $d$ is the cloud distance, $\mu$ is the mean molecular weight corrected for the helium abundance, $\beta_K \simeq 1.67 \times 10^{22} \mbox{ cm}^{-2} \mbox{ mag}^{-1}$ is the ratio $\bigl[ N(\mathrm{H\textsc{i}}) + 2N(\mathrm{H}_2) \bigr] / A_K$ (\citealp{Savage1979}; see also \citealp{Lilley1955,Bohlin1978}), and the integral is evaluated over the whole field $\Omega$, the area of the survey defined in Figure~\ref{fig:5} by a solid line (an area that avoids as much as possible contamination from the red stellar population of the Sagittarius Dwarf). This area is defined in table~\ref{tab:region}.

\begin{table}[h!]
  \centering
  \begin{tabular}{lc}
Corona: & Corona 1 $\cup$ Corona 2  \\
\hline
Corona 1: & $282^\circ < \alpha < 292^\circ$, $-35^\circ < \delta < -41^\circ$ \\ [0.05cm]
Corona 2: & $292^\circ < \alpha < 306^\circ$, $-25^\circ < \delta < -35^\circ$ 
  \end{tabular}
  \caption{The region considered for the mass determination of the cloud (see Figure~\ref{fig:5}.)}
\label{tab:region}
\end{table}

The cloud area considered in this paper is larger than the one used to calculate the mass of this cloud in \citet{Lada2010}, explaining why the mass derived in this paper is also higher.  Assuming a standard cloud composition ($63\%$ hydrogen, $36\%$ helium, and $1\%$ dust), we find $\mu = 1.37$.  Note that, to a certain extent, the ``total'' mass of a cloud is a ill defined quantity, because the boundaries of a cloud are somewhat arbitrary.  In addition, if we go to very small extinction values, relatively small calibration errors on the extinction zeropoint can have a very large impact on the mass.  For this reason, it is more sensible to consider the mass of the cloud above some threshold (for example $A_K > 0.1 \mbox{ mag}$; see Table~\ref{tab:4}).

\citet{Lada2010} found that the star formation rate (SFR) in molecular clouds is linearly proportional to the cloud mass above an extinction threshold of A$_{K} \sim 0.8$ mag.  The Corona cloud region with extinction larger than A$_K > 0.8 \mbox{ mag}$ accounts for $\sim 3.6\%$ of the total integrated mass of the complex. This small percentage has, besides its physical significance for the structure of the cloud, a practical importance regarding the robustness of the extinction map presented in this paper, namely, that we do not expect any significant underestimation in the cloud mass due to unresolved dense cores in our map.

\begin{table}[h]
  \centering
  \begin{tabular}{lcccc}
    Cloud    & Distance &
    \multicolumn{3}{c}{Mass} \\ [0.05cm]
    \cline{3-5}
             &          & Total & $A_K > 0.1$ & $A_K > 0.2$ \\
    \hline
Corona  & $130 \mbox{ pc}$ & $\phantom{0}7\,060 \mbox{ M}_\odot$
    & $\phantom{0} 4\,200 \mbox{ M}_\odot$ 
    & $\phantom{0}950 \mbox{ M}_\odot$ \\ [0.05cm]
  \end{tabular}
  \caption{The mass estimate for the Corona Australis molecular
    complex. }
  \label{tab:4}
\end{table}

\subsection{Fraction of mass below A$_V\sim1$ mag}
One of the findings of the present work is that the Corona complex is about three times larger as revealed by previous molecular lines and dust emission surveys,  due to the detection of a diffuse and extended \textit{``tail''} with an average column density below A$_V \sim1$ mag.   Integrating the dust extinction measurements over these two cloud components we find that $\sim65\%$ of the mass of the Corona complex lies below A$_V \sim1$ mag. It is remarkable that about two thirds of the projected area of the Corona complex ($\sim50^{\circ 2}$) lie beneath the critical density for the formation of CO \citep[A$_V \sim1.6$ mag, e.g.][]{Alves1999,Pineda2010}.

It was early realized, with the advent of the IRAS all-sky survey, that dust column density could trace cloud material beyond the boundaries of the CO emission \citep[e.g.][]{de-Vries1987,Heiles1988,Blitz1990}. Theoretically it has been argued that the formation of CO occurs deeper into the cloud than the formation of H$_2$, because of more effective self-shielding, \citep[e.g.][]{van-Dishoeck1988}, implying the existence of a lower column density region in a molecular cloud with little or no CO. This ``H$_2$-only'' region will escape detection both from CO emission and HI emission, and is also known as ``dark gas'' \citep{Grenier2005}, as H$_2$ is essentially undetectable for the relevant temperatures in these clouds (10--100 K). An important fraction of the gas in the \textit{``tail''} region is likely to be ``dark gas'', although future CO and HI observations are needed to confirm this prediction. In fact, the Corona complex is an ideal laboratory for the study of the transitions HI--H$_2$ and C$^+$--CO in molecular clouds because it is a relatively simple and clean environment in comparison to other complexes closer to the Galactic plane.

\section{Column density PDF: gaussian, log-normal, or power-law?}
\label{sec:column-dens-prob}

\begin{figure*}[!tbp]
  \begin{center}
    \includegraphics[width=13cm]{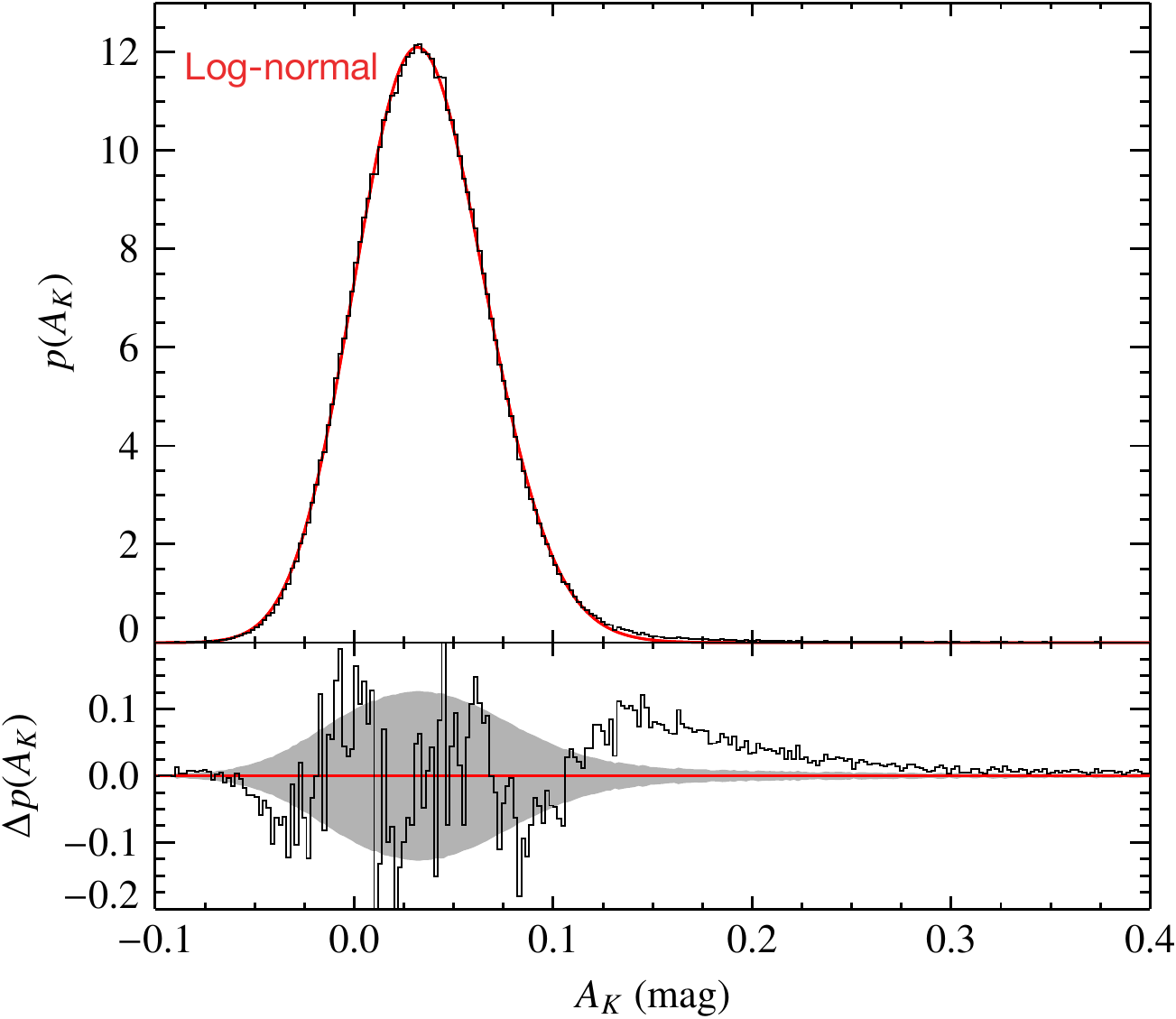}
    \caption{The PDF of column density for the Corona Australis cloud.  The red solid curve represents the best-fit of a single log-normal distribution to the data. The residuals of the fit are presented in the lower panel. The grey area in the residual plot delimits what would be expected from poisson noise. Note that the residuals are not poissonian but exhibit a clear pattern. The excess over the log-normal fit on the high extinction tail is present in a similar manner to the results for the clouds in Paper~III and IV, and the results of \citet{Kainulainen2009c} and \citet{Froebrich2010}.}
    \label{fig:11}
  \end{center}
\end{figure*}

The probability density function (PDF) for the volume density in molecular clouds is expected to be log-normal for isothermal turbulent flows \citep[e.g.][]{Vazquez-Semadeni1994,Padoan1997,Scalo1998,Passot1998,2001ApJ...546..980O,2007ApJ...665..416K,2007A&A...465..445H,2011MNRAS.416.1436B,2011ApJ...727L..20K,2012ApJ...761..156F,2013ApJ...763...51F}, and under certain assumptions the PDF of the column density is also expected to follow a log-normal distribution \citep{2001ApJ...546..980O,Vazquez-Semadeni2001}. Recently, two papers called for non-lognormal column density fits to the PDFs of interstellar clouds. \citet{2010A&A...512A..81F} found a strong dependence of the density PDF on the type of turbulent forcing and that different observed regions show evidence of different mixtures of compressive and solenoidal forcing. They find that a skewed log-normal PDF fit provides a better representation of the column density derived from numerical simulations, although still missing the high-density tail of the PDF obtained for compressive forcing. In a related  recent paper \citet{2013MNRAS.430.1880H} argued for a physically motivated fitting function for density PDFs in turbulent, ideal gas. This function provides a better fit to simulations than both the pure log-normal and the skewed log-normal. As a caveat, \citet{Tassis2010}  used simulations to demonstrate that log-normal column density distributions are generic features of diverse model clouds, and should not be interpreted as being a consequence of supersonic turbulence.

The comparison between the predictions from supersonic flow simulations with observational data have given mixed results.  In a recent investigation of the Perseus cloud \citet{Goodman2009} found no obvious relation between Mach number and normalized column density variance, raising questions on the suggested relation between Mach number and the width of a log-normal column density PDF \citep[e.g.][]{Padoan1997,2001ApJ...546..980O}. On the other hand, using 2MASS NIR extinction maps, \citet{Kainulainen2009c} have characterized the shape of the column density PDFs in nearby molecular clouds and found that although the peaks of the PDFs were generally consistent with log-normal distributions, there were systematic excess “wings” at higher column densities for clouds currently forming stars (including Corona Australis).  Using a similar approach, a similar cloud sample, and the same data base (2MASS), \citet{2007MNRAS.378.1447F,Froebrich2010} also found that some clouds show an excess of column density compared to a log-normal distribution at higher column densities, although they did not find a significant correlation with star formation.  Recently, \citet{2013ApJ...766L..17S} also found a power-law excess over a log-normal column density PDF for Orion B from Herschel data. Both \citet{Kainulainen2009c} and \citet{Froebrich2010} suggested that the observed excess material over the log-normal PDF represents the cloud material decoupled from the general turbulent field and dominated by gravity. \citet{2013ApJ...766L..17S}, also argues that the tail is related to star formation, but stresses that statistical density fluctuations, intermittency, and magnetic fields can also cause the observed excess.

\begin{figure*}[!tbp]
  \begin{center}
    \includegraphics[width=13cm]{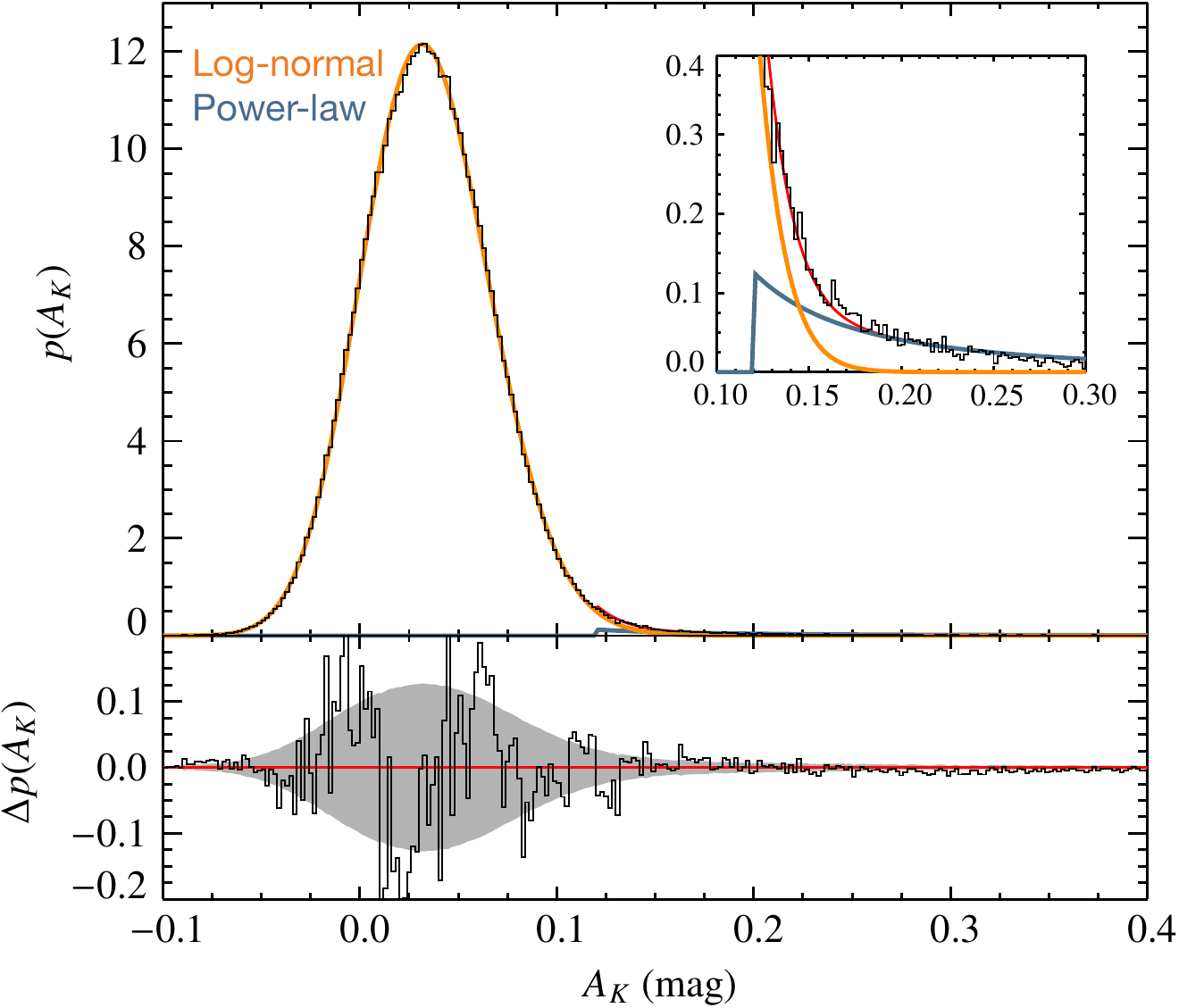}
    \caption{Column density PDF two component fit of Corona: the orange and blue curves correspond to a log-normal and a power-law fit to the observed PDF. The red solid curve represents the sum of the orange and blue curves (below $\sim0.12$ mag, the red curve coincides with the orange curve). The fit of these two functions to the entire cloud produced better residuals (grey area) than for the case of a single log-normal, but a clear pattern in the residuals at lower column densities is still present.
}
    \label{fig:32d}
  \end{center}
\end{figure*}

\begin{figure*}[!tbp]
  \begin{center}
    \includegraphics[width=13cm]{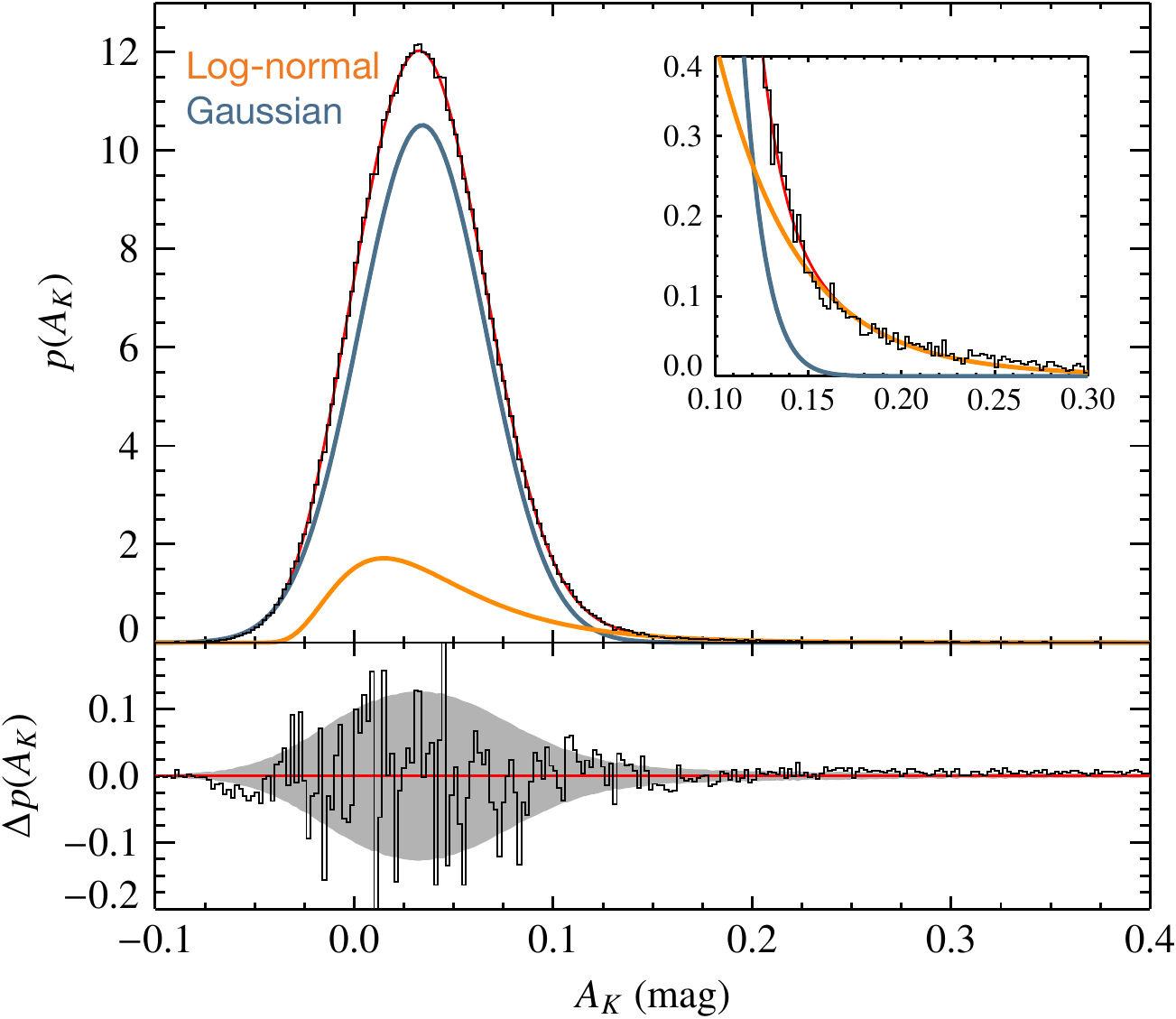}
    \caption{Column density PDF two component fit of Corona: the orange and blue curves corresponds to a log-normal and a gaussian simultaneous fits.  The red solid curve represents the sum of the orange and blue curves. The fit of these two functions to the entire cloud produced the best fit, with residuals consistent with the expected uncorrelated errors (grey area).
}
    \label{fig:11c}
  \end{center}
\end{figure*}

\subsection{A single log-normal PDF}
In this section we take a closer look at the column density PDF of the Corona Australis cloud. Figure~\ref{fig:11} shows the probability density distribution (PDF) of column density for the entire area of the cloud as defined in Figure~\ref{fig:5}. As in our previous papers of this series, we find a significant number of column density estimates with negative values. This is due to uncertainties in the column density measurements, which naturally broadens the intrinsic distribution and adds a fraction of negative measurements. Note, however, that the amount of negative pixels observed is compatible with the typical error on our extinction maps, which is of the order of 0.02 mag.  Also shown in Figure~\ref{fig:11} is a best fit of a single log-normal function\footnote{Note that the functional form used here differs, in the definition of $\sigma_\mathrm{ln}$, with respect to the form used in the papers II and III, but is the same as in paper IV.} to the data (red solid curve), of the form:

 \begin{equation}
  \label{eq:4}
  h(A_K) = \frac{a}{A_K - A_0} 
  \exp\left[- \frac{\bigl(\ln (A_K - A_0) - \ln A_1 \bigr)^2}%
    {2 \sigma_\mathrm{ln}^2} \right] \; .
\end{equation}

\noindent where $a$ is the normalization factor, $A_0$ is the offset, the mean is given by $A_1\times e^{\sigma_\mathrm{ln}^2/2}+A_0$, and the median of the distribution is $A_1+A_0$. The offset  $A_0$ is introduced to allow the fit to explore negative values of A$_K$. The fit parameters are listed in Table~\ref{table:1}. The bottom panel shows the residuals of the fit, and the expected 1$\sigma$ error (grey area). Examination of the residuals shows two significant features that deviate from the expected errors. First, the residuals display a clear extended excess in the high-extinction wing of the PDF. Second, although the amplitude of the residuals is consistent with expectations in the core of the PDF, the residuals exhibit a \textit{systematic correlated pattern} of noise that deviates from the expectation of uncorrelated errors. The clouds in Paper~III and Paper~IV (Perseus, Taurus, California, Orion, Mon R2, Rosette, and Canis Major) all displayed a similar pattern in the residuals as seen in Figure~\ref{fig:11}, suggesting that this is a general problem affecting all log-normal fitting involving the core of the PDF, and not particular to this cloud.

It is clear from Figure~\ref{fig:11}  that a single function cannot account sufficiently for the observed PDF. There is, nevertheless, enough motivation to try to perform a two-component fit to the cloud PDF, because of the apparent power-law tail at high column densities over a log-normal PDF discussed in the literature, to account for noise, or to account for the possible the presence of unresolved spatial variations in the PDF. Recently, regional variations in the column density PDF within a single cloud were found \citep{2008ApJ...679..481P,2012MNRAS.423.2579B,2012A&A...540L..11S}, that suggest that superposition of different PDF components is probably common. 
In the following paragraphs we will investigate two cases of a two-component fit, namely: 1) a log-normal plus a power-law tail,  2) a log-normal and a gaussian, and 3) a gaussian + a power-law, to investigate the impact of the errors on these fits. We performed our fits by simultaneously adjusting all fit parameters for all distributions: in other words, we did not fit separately different parts of the PDF using different functional forms, but rather we fit the entire range with the \textit{sum} of all functional forms selected (log-normal + power law, log-normal+gaussian, and gaussian + power law).

\subsection{Log-normal + power-law}
Regarding the first case, a log-normal plus a power-law tail, this is the most accepted case in the literature \citep{Kainulainen2009c,Froebrich2010,2013ApJ...766L..17S}.  We performed a simultaneous fit of a log-normal and a power-law to the data and the results are presented in Figure~\ref{fig:32d}. The parameters for this fit are listed in Table~\ref{table:1}. One can see both from the Figure and the results of the fit that this is a better fit than a single log-normal as discussed in Figure~\ref{fig:11}. While the addition of the power-law component improves the residuals at higher column densities, it is clear that it could not account for the systematic correlated pattern of noise in the residuals at the lower extinctions in the vicinity the peak of the PDF.

\begin{figure*}[!tbp]
  \begin{center}
    \includegraphics[width=13cm]{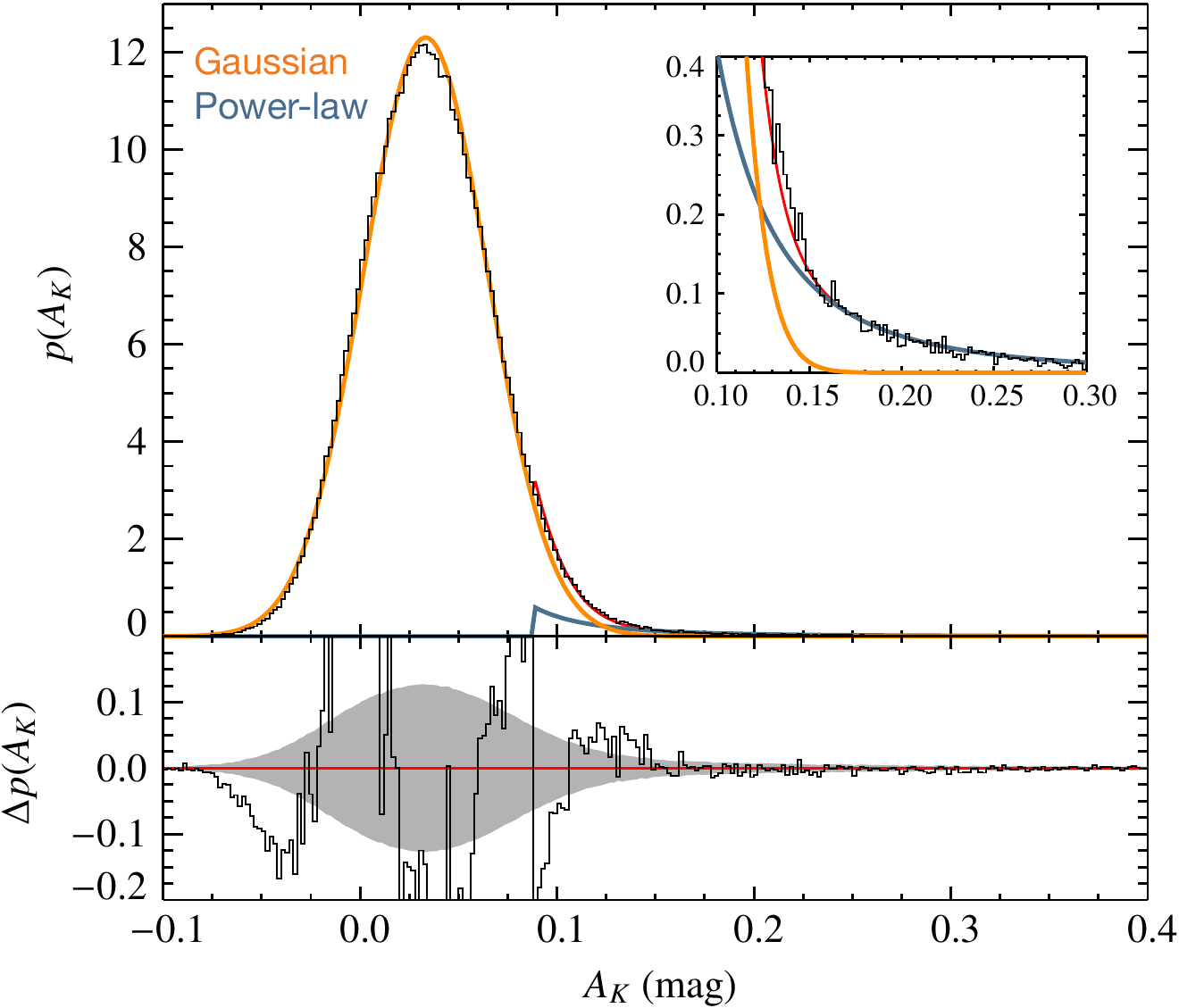}
    \caption{Column density PDF two component fit of Corona: the orange and blue curves corresponds to a gaussian and a power-law simultaneous fits.  The red solid curve represents the sum of the orange and blue curves (below $\sim0.09$ mag, the red curve coincides with the orange curve). The fit of these two functions to the entire cloud produced the lowest $\chi^2$, although presenting residuals exhibiting a systematic correlated pattern at the low-end of the distribution.
}
    \label{fig:32e}
  \end{center}
\end{figure*}

\begin{table*}
\caption{PDF fitting parameters for the different cases discussed (in magnitudes of  A$_K$)}             
\label{table:1}      
\centering          
\begin{tabular}{l c c c c c c}     
\hline\hline       
Case & $\mu$ & $\sigma$ & Offset & Ak cut & Exponent & $\chi^2$ \\ 
\hline   
&&&&&& \\                 
   single log-normal & 0.589 & 0.055 &  $-0.555$ & & & 25508 \\  
&&&&&& \\
\hline
&&&&&& \\
   log-normal & 0.535 & 0.061 & $-0.501$  &  &  &  4870\\
   power-law &           &          &  & 0.121& $-2.22$ & 4870\\
&&&&&& \\
\hline
&&&&&& \\
   log-normal & 0.088 & 0.474 & $-0.055$ & & & 1122\\
   gaussian & 0.034 & 0.032   & & & & 1122\\
&&&&&& \\
   \hline     
&&&&&& \\
   gaussian & 0.033 & 0.032 & & & & 911\\
   power-law & & & & 0.08 & $-3.14$ & 911\\
&&&&&& \\
   \hline                            
\end{tabular}
\end{table*}

\subsection{Log-normal + gaussian}

The presence of negative values of extinction in the observed PDF of the Corona cloud indicates that noise may contribute a non-negligible signal to the overall PDF. To estimate the magnitude of a possible noise component to the cloud PDF we constructed the PDF of pixels off the main cloud. It peaks close to zero extinction and has a gaussian shape as would be expected for measurements dominated by noise. Moreover its FWHM is significant ($\sim0.038$ mag) and we conclude that such a noise profile may represent a major component of the Corona PDF. Therefore we performed a simultaneous fit of a gaussian and a log-normal function. 

The results are shown in Figure~\ref{fig:11c} and the parameters of the fit are listed in Table~\ref{table:1}. This is clearly a better fit than the previous case, as both the excess at high column densities and the systematic correlated pattern of noise in the residuals have vanished, as well as the lower $\chi^2$. The interpretation for this case seems straightforward: the inherent errors in the extinction measurements (blue curve) dominate the core of the observed column density PDF and are the main cause for the correlated noise pattern in the residuals seen in the earlier fits (Figures~\ref{fig:11} and~\ref{fig:32d}). This has important consequences for interpreting the PDFs derived from infrared extinction measurements of clouds like the  Corona. The cores of such PDFs are dominated by noise. The observed cloud PDF is thus the convolution of the noise with the underlying or intrinsic PDF of the cloud, which at infrared extinctions below approximately 0.1 mag is relatively weak compared to the noise. This may make it very difficult to infer the true nature of the intrinsic cloud PDF at levels of extinction below A$_K \approx 0.15$ magnitudes. Thus, although a gaussian + log-normal function provides an excellent fit to the observed PDF of the Corona cloud, it is not clear that this conclusively indicates that the intrinsic cloud PDF is in reality characterized by the specific parameters derived from the fit or even if it is a log-normal in shape.

\subsection{Gaussian + power-law}

We finally performed a simultaneous fit with a gaussian and a power-law. The motivation for this is two folded: 1) the core of the PDF appears to first order Gaussian, and 2) completeness, as this is the last plausible case left. The results are shown in Figure~\ref{fig:32e} and the parameters of the fit are listed in Table~\ref{table:1}. Surprisingly, this fit gave the lowest $\chi^2$ of all cases discussed, suggesting that a simple power-law could be possible description for the intrinsic PDF of the cloud for A$_K >0.15$ mag, in fact better than the log-normal discussed in the previous case. Unlike the previous case, the residuals at low column density show a correlated pattern of noise.

The last two cases, namely ``gaussian + log-normal'' and ``gaussian + power-law'' are likely the best candidates to explain the observations. With respect to the residuals, the  ``gaussian + log-normal'' seem to give the best overall fit, however, in terms of $\chi^2$, the ``gaussian + power-law'' is slightly preferred, and in particular fits best for  A$_k$ above $\sim0.1$ mag. This suggests that a power-law with exponent $-3$ could be the best representation of the PDF for the molecular part of the complex.  On the other hand, the ``gaussian + power-law'' fails to reproduce the residuals at the low column densities. These considerations make it difficult to decide which model best describes the underlying structure of the cloud. 

We conclude that with current data the intrinsic PDF of the Corona Australis cloud is not precisely determinable over the full range of column densities, in particular at the low end where the noise dominates the PDF. This problem is unlikely unique to Corona and should apply to all column density maps affected by even moderate noise (average S/N $\sim5$ ). Given the prevalence of the features discussed above for clouds in Paper~III and Paper~IV (Perseus, Taurus, California, Orion, Mon R2, Rosette, and Canis Major), one expects the case of Corona to be hardly unique, suggesting that the parameters derived from the observed PDFs currently discussed in the literature may be flawed. A way forward would be to improve the knowledge of the low column density regions in clouds. This might be possible with derived column density maps from ESA's Planck satellite.

Finally, the diffuse component identified in Section~\ref{sec:statistical-analysis} for the Corona cloud may not be unique, although this would need to be confirmed in the future.  While we can only speculate about the origin or the role of these diffuse envelopes in the formation and evolution of molecular clouds (e.g., are they being ablated from the denser cloud regions by stellar feedback from nearby massive stars, or do they indicate molecular cloud formation?), we can be sure that they are an important component in the structure, size, and mass budget of the Corona cloud complex.  In Corona we have found what is likely to be the best case for an isolated molecular cloud with a diffuse (atomic?) envelope. Because of this, the Corona Australis cloud complex might turn out to be our best  laboratory to study the relation between the atomic and molecular components in clouds as well as understanding the role of turbulence and external feedback in cloud and star formation.

\section{Conclusions}
The following items summarize the main results presented in this paper:
\begin{enumerate}

\item We measured the extinction over an area of $\sim870$ square degrees centered Corona Australis cloud. The extinction map, obtained with the \textsc{Nicer} and \textsc{Nicest} algorithms, has a resolution of 3 arcmin and an average 1-$\sigma$ detection level of A$_K\sim0.02$ mag.

\item As seen in projection, the Corona Australis cloud consists of  a 45 pc curved complex of filamentary structure.  While the star forming Western-end (the \textit{``head''}) has a peak column density above 10$^{23}$ cm$^{-2}$ (or A$_K>5$ mag), the sensitivity of the extinction map allows us to map the extended and diffuse Eastern-end \textit{``tail''} of the cloud down to a mean column density of $\sim10^{21}$ cm$^{-2}$ (or A$_K\sim0.05$ mag). About two thirds of the mass of the complex (65\%) lies beneath A$_V\sim1$ mag.

\item We find that the PDF of the cloud cannot be described by a single log-normal function, similar to prior studies. We show that at low column densities near the peak of the observed PDF, both the amplitude and shape of the PDF, are dominated by noise in the extinction measurements making it impractical to derive the intrinsic cloud PDF  below  A$_K < 0.15$ mag. Above A$_K \sim 0.15$ mag, the molecular component of the cloud, the PDF appears to be best described by a power-law with index $-3$, but could also described as the tail of a broad and relatively low amplitude, log-normal PDF that peaks at very low column densities.

\end{enumerate}

\noindent
Multi-population stellar backgrounds, like the one towards the Sagittarius Dwarf Galaxy, present a new challenge to NIR extinction mapping techniques. This challenge needs to be addressed for the exploitation of upcoming deep NIR surveys.

\begin{acknowledgements}
It is a pleasure to acknowledge discussions with Paolo Padoan, Jouni Kainulainen, and Alyssa Goodman. We also want to thank an anonymous referee for comments that subtantially improved the paper. This research made use of Astropy, a community-developed core Python package for Astronomy \citep{2013A&A...558A..33A}. This publication is supported by the Austrian Science Fund (FWF). This research has made use of the 2MASS archive, provided by NASA/IPAC Infrared Science Archive, which is operated by the Jet Propulsion Laboratory, California Institute of Technology, under contract with the National Aeronautics and Space Administration.  CJL acknowledges support from NASA ORIGINS Grant NAG 5-13041.
\end{acknowledgements}

\bibliographystyle{aa} 
\bibliography{/Users/jalves/Dropbox/Research/zmy}

\begin{thebibliography}{55}
\expandafter\ifx\csname natexlab\endcsname\relax\def\natexlab#1{#1}\fi

\bibitem[{{Alves} {et~al.}(1999){Alves}, {Lada}, \& {Lada}}]{Alves1999}
{Alves}, J., {Lada}, C.~J., \& {Lada}, E.~A. 1999, \apj, 515, 265

\bibitem[{{Astropy Collaboration} {et~al.}(2013){Astropy Collaboration},
  {Robitaille}, {Tollerud}, {Greenfield}, {Droettboom}, {Bray}, {Aldcroft},
  {Davis}, {Ginsburg}, {Price-Whelan}, {Kerzendorf}, {Conley}, {Crighton},
  {Barbary}, {Muna}, {Ferguson}, {Grollier}, {Parikh}, {Nair}, {Unther},
  {Deil}, {Woillez}, {Conseil}, {Kramer}, {Turner}, {Singer}, {Fox}, {Weaver},
  {Zabalza}, {Edwards}, {Azalee Bostroem}, {Burke}, {Casey}, {Crawford},
  {Dencheva}, {Ely}, {Jenness}, {Labrie}, {Lim}, {Pierfederici}, {Pontzen},
  {Ptak}, {Refsdal}, {Servillat}, \& {Streicher}}]{2013A&A...558A..33A}
{Astropy Collaboration}, {Robitaille}, T.~P., {Tollerud}, E.~J., {et~al.} 2013,
  \aap, 558, A33

\bibitem[{{Ballesteros-Paredes} {et~al.}(2011){Ballesteros-Paredes},
  {V{\'a}zquez-Semadeni}, {Gazol}, {Hartmann}, {Heitsch}, \&
  {Col{\'{\i}}n}}]{2011MNRAS.416.1436B}
{Ballesteros-Paredes}, J., {V{\'a}zquez-Semadeni}, E., {Gazol}, A., {et~al.}
  2011, \mnras, 416, 1436

\bibitem[{{Beaumont} {et~al.}(2012){Beaumont}, {Goodman}, {Alves}, {Lombardi},
  {Rom{\'a}n-Z{\'u}{\~n}iga}, {Kauffmann}, \& {Lada}}]{2012MNRAS.423.2579B}
{Beaumont}, C.~N., {Goodman}, A.~A., {Alves}, J.~F., {et~al.} 2012, \mnras,
  423, 2579

\bibitem[{{Blitz} {et~al.}(1990){Blitz}, {Bazell}, \& {Desert}}]{Blitz1990}
{Blitz}, L., {Bazell}, D., \& {Desert}, F.~X. 1990, \apjl, 352, L13

\bibitem[{{Bohlin} {et~al.}(1978){Bohlin}, {Savage}, \& {Drake}}]{Bohlin1978}
{Bohlin}, R.~C., {Savage}, B.~D., \& {Drake}, J.~F. 1978, \apj, 224, 132

\bibitem[{{Casey} {et~al.}(1998){Casey}, {Mathieu}, {Vaz}, {Andersen}, \&
  {Suntzeff}}]{Casey1998}
{Casey}, B.~W., {Mathieu}, R.~D., {Vaz}, L.~P.~R., {Andersen}, J., \&
  {Suntzeff}, N.~B. 1998, \aj, 115, 1617

\bibitem[{{Chini} {et~al.}(2003){Chini}, {K{\"a}mpgen}, {Reipurth}, {Albrecht},
  {Kreysa}, {Lemke}, {Nielbock}, {Reichertz}, {Sievers}, \&
  {Zylka}}]{Chini2003}
{Chini}, R., {K{\"a}mpgen}, K., {Reipurth}, B., {et~al.} 2003, \aap, 409, 235

\bibitem[{{Dame} {et~al.}(2001){Dame}, {Hartmann}, \& {Thaddeus}}]{Dame2001}
{Dame}, T.~M., {Hartmann}, D., \& {Thaddeus}, P. 2001, \apj, 547, 792

\bibitem[{{de Vries} {et~al.}(1987){de Vries}, {Thaddeus}, \&
  {Heithausen}}]{de-Vries1987}
{de Vries}, H.~W., {Thaddeus}, P., \& {Heithausen}, A. 1987, \apj, 319, 723

\bibitem[{{Federrath} \& {Klessen}(2012)}]{2012ApJ...761..156F}
{Federrath}, C. \& {Klessen}, R.~S. 2012, \apj, 761, 156

\bibitem[{{Federrath} \& {Klessen}(2013)}]{2013ApJ...763...51F}
{Federrath}, C. \& {Klessen}, R.~S. 2013, \apj, 763, 51

\bibitem[{{Federrath} {et~al.}(2010){Federrath}, {Roman-Duval}, {Klessen},
  {Schmidt}, \& {Mac Low}}]{2010A&A...512A..81F}
{Federrath}, C., {Roman-Duval}, J., {Klessen}, R.~S., {Schmidt}, W., \& {Mac
  Low}, M.-M. 2010, \aap, 512, A81

\bibitem[{{Forbrich} \& {Preibisch}(2007)}]{Forbrich2007}
{Forbrich}, J. \& {Preibisch}, T. 2007, \aap, 475, 959

\bibitem[{{Froebrich} {et~al.}(2007){Froebrich}, {Murphy}, {Smith}, {Walsh}, \&
  {Del Burgo}}]{2007MNRAS.378.1447F}
{Froebrich}, D., {Murphy}, G.~C., {Smith}, M.~D., {Walsh}, J., \& {Del Burgo},
  C. 2007, \mnras, 378, 1447

\bibitem[{{Froebrich} \& {Rowles}(2010)}]{Froebrich2010}
{Froebrich}, D. \& {Rowles}, J. 2010, \mnras, 406, 1350

\bibitem[{{Goodman} {et~al.}(2009){Goodman}, {Pineda}, \&
  {Schnee}}]{Goodman2009}
{Goodman}, A.~A., {Pineda}, J.~E., \& {Schnee}, S.~L. 2009, \apj, 692, 91

\bibitem[{{Grenier} {et~al.}(2005){Grenier}, {Casandjian}, \&
  {Terrier}}]{Grenier2005}
{Grenier}, I.~A., {Casandjian}, J.-M., \& {Terrier}, R. 2005, Science, 307,
  1292

\bibitem[{{Harju} {et~al.}(1993){Harju}, {Haikala}, {Mattila}, {Mauersberger},
  {Booth}, \& {Nordh}}]{Harju1993}
{Harju}, J., {Haikala}, L.~K., {Mattila}, K., {et~al.} 1993, \aap, 278, 569

\bibitem[{{Heiles} {et~al.}(1988){Heiles}, {Reach}, \& {Koo}}]{Heiles1988}
{Heiles}, C., {Reach}, W.~T., \& {Koo}, B.-C. 1988, \apj, 332, 313

\bibitem[{{Hennebelle} {et~al.}(2007){Hennebelle}, {Audit}, \&
  {Miville-Desch{\^e}nes}}]{2007A&A...465..445H}
{Hennebelle}, P., {Audit}, E., \& {Miville-Desch{\^e}nes}, M.-A. 2007, \aap,
  465, 445

\bibitem[{{Hopkins}(2013)}]{2013MNRAS.430.1880H}
{Hopkins}, P.~F. 2013, \mnras, 430, 1880

\bibitem[{{Indebetouw} {et~al.}(2005){Indebetouw}, {Mathis}, {Babler}, {Meade},
  {Watson}, {Whitney}, {Wolff}, {Wolfire}, {Cohen}, {Bania}, {Benjamin},
  {Clemens}, {Dickey}, {Jackson}, {Kobulnicky}, {Marston}, {Mercer},
  {Stauffer}, {Stolovy}, \& {Churchwell}}]{Indebetouw2005}
{Indebetouw}, R., {Mathis}, J.~S., {Babler}, B.~L., {et~al.} 2005, \apj, 619,
  931

\bibitem[{{Kainulainen} {et~al.}(2009){Kainulainen}, {Beuther}, {Henning}, \&
  {Plume}}]{Kainulainen2009c}
{Kainulainen}, J., {Beuther}, H., {Henning}, T., \& {Plume}, R. 2009, \aap,
  508, L35

\bibitem[{{Kleinmann} {et~al.}(1994){Kleinmann}, {Lysaght}, {Pughe},
  {Schneider}, {Skrutskie}, {Weinberg}, {Price}, {Matthews}, {Soifer}, \&
  {Huchra}}]{Kleinmann1994}
{Kleinmann}, S.~G., {Lysaght}, M.~G., {Pughe}, W.~L., {et~al.} 1994,
  Experimental Astronomy, 3, 65

\bibitem[{{Kritsuk} {et~al.}(2007){Kritsuk}, {Norman}, {Padoan}, \&
  {Wagner}}]{2007ApJ...665..416K}
{Kritsuk}, A.~G., {Norman}, M.~L., {Padoan}, P., \& {Wagner}, R. 2007, \apj,
  665, 416

\bibitem[{{Kritsuk} {et~al.}(2011){Kritsuk}, {Norman}, \&
  {Wagner}}]{2011ApJ...727L..20K}
{Kritsuk}, A.~G., {Norman}, M.~L., \& {Wagner}, R. 2011, \apjl, 727, L20

\bibitem[{{Lada} {et~al.}(1994){Lada}, {Lada}, {Clemens}, \&
  {Bally}}]{Lada1994}
{Lada}, C.~J., {Lada}, E.~A., {Clemens}, D.~P., \& {Bally}, J. 1994, \apj, 429,
  694

\bibitem[{{Lada} {et~al.}(2010){Lada}, {Lombardi}, \& {Alves}}]{Lada2010}
{Lada}, C.~J., {Lombardi}, M., \& {Alves}, J.~F. 2010, \apj, 724, 687

\bibitem[{{Lilley}(1955)}]{Lilley1955}
{Lilley}, A.~E. 1955, \apj, 121, 559

\bibitem[{{Lombardi}(2009)}]{Lombardi2009}
{Lombardi}, M. 2009, \aap, 493, 735

\bibitem[{{Lombardi} \& {Alves}(2001)}]{Lombardi2001}
{Lombardi}, M. \& {Alves}, J. 2001, \aap, 377, 1023

\bibitem[{{Lombardi} {et~al.}(2006){Lombardi}, {Alves}, \&
  {Lada}}]{Lombardi2006}
{Lombardi}, M., {Alves}, J., \& {Lada}, C.~J. 2006, \aap, 454, 781

\bibitem[{{Lombardi} {et~al.}(2011){Lombardi}, {Alves}, \&
  {Lada}}]{2011A&A...535A..16L}
{Lombardi}, M., {Alves}, J., \& {Lada}, C.~J. 2011, \aap, 535, A16

\bibitem[{{Lombardi} {et~al.}(2008){Lombardi}, {Lada}, \&
  {Alves}}]{Lombardi2008}
{Lombardi}, M., {Lada}, C.~J., \& {Alves}, J. 2008, \aap, 489, 143

\bibitem[{{Lombardi} {et~al.}(2010){Lombardi}, {Lada}, \&
  {Alves}}]{Lombardi2010}
{Lombardi}, M., {Lada}, C.~J., \& {Alves}, J. 2010, \aap, 512, 67

\bibitem[{{Majewski} {et~al.}(2003){Majewski}, {Skrutskie}, {Weinberg}, \&
  {Ostheimer}}]{Majewski2003}
{Majewski}, S.~R., {Skrutskie}, M.~F., {Weinberg}, M.~D., \& {Ostheimer}, J.~C.
  2003, \apj, 599, 1082

\bibitem[{{Neuh{\"a}user} \& {Forbrich}(2008)}]{Neuhauser2008}
{Neuh{\"a}user}, R. \& {Forbrich}, J. 2008, {The Corona Australis Star Forming
  Region (Handbook of Star Forming Regions, Volume II)}, ed. {Reipurth, B.},
  735

\bibitem[{{Ostriker} {et~al.}(2001){Ostriker}, {Stone}, \&
  {Gammie}}]{2001ApJ...546..980O}
{Ostriker}, E.~C., {Stone}, J.~M., \& {Gammie}, C.~F. 2001, \apj, 546, 980

\bibitem[{{Padoan} {et~al.}(1997){Padoan}, {Nordlund}, \& {Jones}}]{Padoan1997}
{Padoan}, P., {Nordlund}, A., \& {Jones}, B.~J.~T. 1997, \mnras, 288, 145

\bibitem[{{Passot} \& {V{\'a}zquez-Semadeni}(1998)}]{Passot1998}
{Passot}, T. \& {V{\'a}zquez-Semadeni}, E. 1998, \pre, 58, 4501

\bibitem[{{Peterson} {et~al.}(2011){Peterson}, {Caratti o Garatti}, {Bourke},
  {Forbrich}, {Gutermuth}, {J{\o}rgensen}, {Allen}, {Patten}, {Dunham},
  {Harvey}, {Mer{\'{\i}}n}, {Chapman}, {Cieza}, {Huard}, {Knez}, {Prager}, \&
  {Evans}}]{Peterson2011}
{Peterson}, D.~E., {Caratti o Garatti}, A., {Bourke}, T.~L., {et~al.} 2011,
  \apjs, 194, 43

\bibitem[{{Pineda} {et~al.}(2008){Pineda}, {Caselli}, \&
  {Goodman}}]{2008ApJ...679..481P}
{Pineda}, J.~E., {Caselli}, P., \& {Goodman}, A.~A. 2008, \apj, 679, 481

\bibitem[{{Pineda} {et~al.}(2010){Pineda}, {Goldsmith}, {Chapman}, {Snell},
  {Li}, {Cambr{\'e}sy}, \& {Brunt}}]{Pineda2010}
{Pineda}, J.~L., {Goldsmith}, P.~F., {Chapman}, N., {et~al.} 2010, \apj, 721,
  686

\bibitem[{{Ridge} {et~al.}(2006){Ridge}, {Di Francesco}, {Kirk}, {Li},
  {Goodman}, {Alves}, {Arce}, {Borkin}, {Caselli}, {Foster}, {Heyer},
  {Johnstone}, {Kosslyn}, {Lombardi}, {Pineda}, {Schnee}, \&
  {Tafalla}}]{Ridge2006}
{Ridge}, N.~A., {Di Francesco}, J., {Kirk}, H., {et~al.} 2006, \aj, 131, 2921

\bibitem[{{Savage} \& {Mathis}(1979)}]{Savage1979}
{Savage}, B.~D. \& {Mathis}, J.~S. 1979, \araa, 17, 73

\bibitem[{{Scalo} {et~al.}(1998){Scalo}, {Vazquez-Semadeni}, {Chappell}, \&
  {Passot}}]{Scalo1998}
{Scalo}, J., {Vazquez-Semadeni}, E., {Chappell}, D., \& {Passot}, T. 1998,
  \apj, 504, 835

\bibitem[{{Schneider} {et~al.}(2013){Schneider}, {Andr{\'e}}, {K{\"o}nyves},
  {Bontemps}, {Motte}, {Federrath}, {Ward-Thompson}, {Arzoumanian},
  {Benedettini}, {Bressert}, {Didelon}, {Di Francesco}, {Griffin}, {Hennemann},
  {Hill}, {Palmeirim}, {Pezzuto}, {Peretto}, {Roy}, {Rygl}, {Spinoglio}, \&
  {White}}]{2013ApJ...766L..17S}
{Schneider}, N., {Andr{\'e}}, P., {K{\"o}nyves}, V., {et~al.} 2013, \apjl, 766,
  L17

\bibitem[{{Schneider} {et~al.}(2011){Schneider}, {Bontemps}, {Simon},
  {Ossenkopf}, {Federrath}, {Klessen}, {Motte}, {Andr{\'e}}, {Stutzki}, \&
  {Brunt}}]{Schneider2011}
{Schneider}, N., {Bontemps}, S., {Simon}, R., {et~al.} 2011, \aap, 529, A1+

\bibitem[{{Schneider} {et~al.}(2012){Schneider}, {Csengeri}, {Hennemann},
  {Motte}, {Didelon}, {Federrath}, {Bontemps}, {Di Francesco}, {Arzoumanian},
  {Minier}, {Andr{\'e}}, {Hill}, {Zavagno}, {Nguyen-Luong}, {Attard},
  {Bernard}, {Elia}, {Fallscheer}, {Griffin}, {Kirk}, {Klessen}, {K{\"o}nyves},
  {Martin}, {Men'shchikov}, {Palmeirim}, {Peretto}, {Pestalozzi}, {Russeil},
  {Sadavoy}, {Sousbie}, {Testi}, {Tremblin}, {Ward-Thompson}, \&
  {White}}]{2012A&A...540L..11S}
{Schneider}, N., {Csengeri}, T., {Hennemann}, M., {et~al.} 2012, \aap, 540, L11

\bibitem[{{Sicilia-Aguilar} {et~al.}(2011){Sicilia-Aguilar}, {Henning},
  {Kainulainen}, \& {Roccatagliata}}]{Sicilia-Aguilar2011}
{Sicilia-Aguilar}, A., {Henning}, T., {Kainulainen}, J., \& {Roccatagliata}, V.
  2011, \apj, 736, 137

\bibitem[{{Tassis} {et~al.}(2010){Tassis}, {Christie}, {Urban}, {Pineda},
  {Mouschovias}, {Yorke}, \& {Martel}}]{Tassis2010}
{Tassis}, K., {Christie}, D.~A., {Urban}, A., {et~al.} 2010, \mnras, 408, 1089

\bibitem[{{van Dishoeck} \& {Black}(1988)}]{van-Dishoeck1988}
{van Dishoeck}, E.~F. \& {Black}, J.~H. 1988, \apj, 334, 771

\bibitem[{{Vazquez-Semadeni}(1994)}]{Vazquez-Semadeni1994}
{Vazquez-Semadeni}, E. 1994, \apj, 423, 681

\bibitem[{{V{\'a}zquez-Semadeni} \&
  {Garc{\'{\i}}a}(2001)}]{Vazquez-Semadeni2001}
{V{\'a}zquez-Semadeni}, E. \& {Garc{\'{\i}}a}, N. 2001, \apj, 557, 727

\end{thebibliography}

\end{document}